\documentclass[fleqn,10pt]{wlscirep}
\title{Experimental violation of Bell inequalities for multi-dimensional systems}

\author[1,a]{Hsin-Pin Lo}
\author[2,a*]{Che-Ming Li}
\author[1,+]{Atsushi Yabushita}
\author[3]{Yueh-Nan Chen}
\author[1]{Chih-Wei Luo}
\author[1,4,5,++]{Takayoshi Kobayashi}

\affil[1]{Department of Electrophysics, National Chiao-Tung University, Hsinchu City 300, Taiwan}
\affil[2]{Department of Engineering Science, National Cheng-Kung University, Tainan City 701, Taiwan}
\affil[3]{Department of Physics and National Center for Theoretical Sciences, National Cheng-Kung University, Tainan City 701, Taiwan}
\affil[4]{Advanced Ultrafast Laser Research Center and Department of
Engineering Science, University of Electro-Communications, Chofu, Tokyo
182-8585, Japan}
\affil[5]{JST, CREST, 5 Sanbancho, Chiyoda-ku, Tokyo 102-0075 Japan}

\affil[a]{These authors contributed equally to this work.}
\affil[*]{cmli@mail.ncku.edu.tw}
\affil[+]{yabushita@mail.nctu.edu.tw}
\affil[++]{kobayashi@ils.uec.ac.jp}

\begin{abstract}
Quantum correlations between spatially separated parts of a $d$-dimensional bipartite system ($d\geq 2$) have no classical analog. Such correlations, also called entanglements, are not only conceptually important, but also have a profound impact on information science. In theory the violation of Bell inequalities based on local realistic theories for $d$-dimensional systems provides evidence of quantum nonlocality. Experimental verification is required to confirm whether a quantum system of extremely large dimension can possess this feature, however it has never been performed for large dimension. Here, we report that Bell inequalities are experimentally violated for bipartite quantum systems of dimensionality $d=16$ with the usual ensembles of polarization-entangled photon pairs. We also estimate that our entanglement source violates Bell inequalities for extremely high dimensionality of $d>4000$. The designed scenario offers a possible new method to investigate the entanglement of multipartite systems of large dimensionality and their application in quantum information processing.
\end{abstract}

\begin{document}

\flushbottom
\maketitle

\thispagestyle{empty}

\section*{Introduction}
Entanglement is one of the remarkable predictions of quantum mechanics \cite{Bell64,Clauser69,Kaszlikowski00,Collins02,Son06}, which was first pointed out by Einstein, Podolsky, and Rosen (EPR) \cite{Einstein35,Schrodinger35}. The predicted correlations cannot be reproduced by any classical theories based on local variables [also called local realistic theories (LRT)]. Bell test experiments \cite{Bell64,Clauser69,Bell87} provide a feasibly approach to verify such correlations in practice. The kernel of these experiments is the Bell inequalities, which are based on the constraint that the correlations exhibited by local variable theories must satisfy. Contradicting Bell inequalities can thus be considered a non-classical indicator of quantum nonlocality. 

One of the most striking works on Bell inequalities for multi-dimensional systems is the reformulation of the Collins, Gisin, Linden, Massar, and Popescu (CGLMP) inequalities \cite{Collins02}. 
As presented by CGLMP in their original scenario, we summarize the three key components in a Bell test experiment. The first part is about the $d$-dimensional system. Each one of the two remote parties, say Alice and Bob, has such a system with dimensionality $d$. Using the Bell-CGLMP inequality, they investigate the quantum nonlocality between the $d$-dimensional systems they have. Since CGLMP concern the impossibility to reproduce quantum correlations with local variable theories, the definition of the $d$-dimensional system then should be generally fitting their purpose. When a $d$-dimensional system is able to be measured by performing properly designed measurements which can provide $d$ possible distinct outcomes, it then can be used to reveal the quantum nonlocality. 
The second part is about the measurements performed by Alice and Bob on their respective systems. In a Bell test experiment, suppose that one of the parties, Alice, can carry out two possible measurements, $A_{1}$ or $A_{2}$, and that the other party, Bob, can carry out two possible measurements,  $B_{1}$ or $B_{2}$, where each measurement has $d$ possible outcomes: $A_{1},A_{2},B_{1},B_{2}=0,1,...,d-1$. 
The last part is that, with sufficient measurement results obtained from the Bell test experiment, one can calculate the joint probabilities $P(A_{a},B_{b})$ for $a,b\in \{1,2\}$. For example, Alice and Bob choose measurements $A_{1}$ and $B_{2}$, respectively. Each time, they measure their respective systems and then have a set of outcome. After sufficient runs of such measurements have been made, a joint probability distribution for $P(A_{1},B_{2})$ can be derived from the experimental outcomes. All the other probabilities $P(A_{a},B_{b})$ used to determine the value of Bell's expression can be measured in the same manner.
The correlations among the outcomes of the measurements exhibited by local-variable theories must obey the generalized Bell inequalities 
\begin{equation}
\big| I_{d,LRT}[P(A_{a},B_{b})] \big| \leq 2.\label{belli}
\end{equation} 
The Bell expression, $I_{d}$, reads:
\begin{eqnarray}
&& I_{d}\equiv\sum_{k=0}^{[d/2]-1}\!\!\big(1-\frac{2k}{d-1}\big)\{+[P(A_{1}=B_{1}+k)+P(B_{1}=A_{2}+k+1)+P(A_{2}=B_{2}+k)+P(B_{2}=A_{1}+k)] \nonumber \\
&&\ \ \ \ \ \ \ \ \ \ \ \ \ \ \ \ \ \ \ \ \ \ \ \ \ \ \ \ \ \ \ \ \ \ \ \ \ \  -[P(A_{1}=B_{1}-k-1)+P(B_{1}=A_{2}-k)+P(A_{2}=B_{2}-k-1)+P(B_{2}=A_{1}-k-1)]\},\label{kernel}
\end{eqnarray}
where $P(A_a=B_b+k)$ denotes the probability of the measurements $A_a$ and $B_b$ with outcomes that differ, modulo $d$, by $k$. Compared with previous significant discoveries \cite{Kaszlikowski00}, this is the first analytical demonstration of Bell inequalities for bipartite quantum systems of arbitrarily high dimensionality. 
   
In addition to the fundamental interest for revealing fascinating aspects of quantum mechanics, Bell inequalities generalized to $d$-dimensional systems and their verified quantum nonlocality are also crucial for a variety of quantum information tasks and engineering protocols \cite{Nielsen00,Pan12}, such as quantum communication \cite{Vaziri02,Groblacher06}, quantum computation \cite{Zhou03}, quantum error correction \cite{Looi08}, and quantum metrology \cite{Lloyd08}. Considerable progress has been made to demonstrate the violation of generalized Bell inequalities. In particular, quantum correlations have recently been revealed \cite{Dada11} by experimentally violating Bell inequalities for bipartite $d$-dimensional quantum systems up to $d=12$. In this seminal demonstration of Bell test experiment, genuine multi-dimensional entanglement~\cite{Mair01} of the orbital angular momentum (OAM) \cite{Beijersbergen92} of photon pairs was generated through spontaneous parametric down-conversion (SPDC) and enhanced by the entanglement concentration. The direct extension of the OAM-entangled photons to show violations of Bell inequalities for $d>12$, however, is restricted by the quality of entanglement due to the intrinsic properties of the SPDC process. The recent trend is toward efficient verifications of high-dimensional OAM entanglement, e.g., by using an entanglement measurement \cite{Romero12} or by developing a novel nonlinear criterion \cite{Krenn14}.  


Here, we use a concept different from that of Ref~\cite{Dada11}. The present concept utilizes multi-dimensional objects composed of multiple pairs of polarization-entangled photon to consider their non-classical bipartite multi-dimensional correlations. From our theoretical method and experimental observations, it was proved for the first time that the non-classical multi-dimensional correlations can readily reveal such joint non-classical effects surely violating the CGLMP inequalities. Our theoretical scheme provides a novel point of view to investigate contradictions of Bell-CGLMP inequalities. Such method explores the states of bipartite $d$-dimensional entangled systems where the state of each system is in a Hilbert space spanned by multiple two-dimensional state vectors. It fully satisfies the needs of the scenario of CGLMP and shows a strict connection between the considered $d$-dimensional entangled states and the settings required for measuring the Bell expression $I_{d}$ in an experiment. Thus, following the scheme, polarization-entangled photon pairs were operated to experimentally observe $d$-dimensional quantum nonlocality up to $d=16$ by performing the Bell test experiment. We also objectively extrapolate the $d$-dimensional quantum nonlocality up to $d=4096$ from measuring the density matrix of the entangled state. Besides the verification of the Bell inequalities, slight modification of the method has also allowed us to estimate the entanglement dimensionality.


The Bell-CGLMP inequality is a general tool that helps to detect quantum nonlocality between two spatially separated $d$-dimensional systems. As described by quantum mechanics for a composite system, the $d$-dimensional state vector of a quantum system can be consisted of two-dimensional states of $N$ subsystems where $d=2^{N}$. In the experiment, we choose entangled photon pairs as such subsystems and consider the dimensionality and quantum nonlocality in the photon polarization degree of freedom (DOF). While the system considered in our experiment is composed of subsystems that are allowed to be created at different times, these subsystems eventually constitute a system with a $d$-dimensional state in the polarization DOF. They can be locally measured to provide possible outcomes for determining the joint probabilities $P(A_{a},B_{b})$. Thus the correlation in the polarization DOF between two ensembles of photons can be completely demonstrated by violating the Bell-CGLMP inequality. Furthermore, since the scheme of CGLMP concerns the correlation in the DOF where the $d$-dimensional system is defined, the temporal DOF then does not play a role in determining both $P(A_{a},B_{b})$ and the Bell expression $I_{d}$ (\ref{kernel}) in our experiment. 

\begin{figure}[t]
\includegraphics[width=9cm]{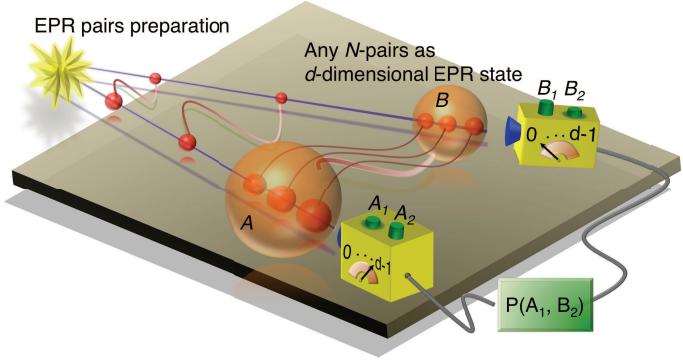}
\caption{Schematic set-up. The verification of quantum nonlocality that contradicts Bell-CGLMP inequalities is implemented with an entanglement source of EPR pairs. The system of ensembles $A$ and $B$ consisting of $N$ EPR pairs is a bipartite $d$-dimensional entangled state $\left|\Phi\right\rangle$ [Eq.~(\ref{Phi})] for $d=2^{N}$. To determine the joint probability $P(A_{a},B_{b})$ the measurement settings are designed to implement unambiguous detections with $d$ possible outcomes.}\label{scheme}
\end{figure}

\section*{Results}

We present in Fig.~\ref{scheme} the theoretical scheme used in our experiments to investigate the quantum nonlocality that contradicts Bell inequalities for systems of large dimensionality. The first step is to prepare the  entanglement of bipartite $d$-dimensional systems. Here, the bipartite system generated for the Bell test experiments is composed of two ensembles of particles. Assuming that each particle is a two-state quantum object with some degree of freedom, the dimension of the Hilbert space of the ensemble consisting of $N$ particles will be $d=2^{N}$. If we assume further that all of the $N$ entangled pairs have the same state of the form $\left|\phi\right\rangle=(\left|0\right\rangle_{A}\otimes\left|0\right\rangle_{B}+\left|1\right\rangle_{A}\otimes\left|1\right\rangle_{B})/\sqrt{2}$, then the state vector reads:
\begin{eqnarray}
\left|\Phi\right\rangle&=&\bigotimes_{m=1}^{N}\frac{1}{\sqrt{2}}(\left|0\right\rangle_{Am}\otimes\left|0\right\rangle_{Bm}+\left|1\right\rangle_{Am}\otimes\left|1\right\rangle_{Bm})\nonumber\\
&=&\frac{1}{\sqrt{d}}\sum_{j=0}^{d-1}\left|j\right\rangle_{A}\otimes\left|j\right\rangle_{B},\label{Phi}
\end{eqnarray}
where $\left|j\right\rangle_{A}=\bigotimes_{m=1}^{N}\left|j_{m}\right\rangle_{Am}$ and $j=\sum_{m=1}^{N}j_{N-m+1}2^{m-1}$ for $j_{m}\in\{0,1\}$. It is a maximally entangled state of two $d$-dimensional systems. In a real experiment, the entangled states that are created are not pure but are mixed states \cite{Pan12}, say, $\rho_{\Phi}(d)$, and, as will be shown presently, the purity of $\rho_{\Phi}(d)$ affects the degree of violation of the Bell inequalities.

\begin{figure}[t]
\includegraphics[width=9cm]{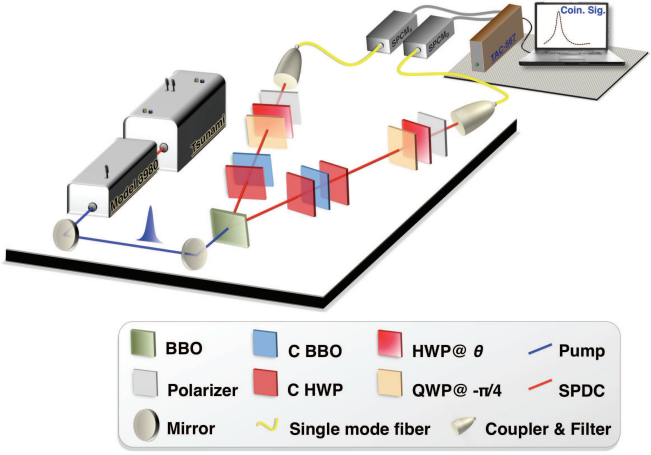}
\caption{Experimental set-up. The ultraviolet pulsed laser (200 mW) is generated by second-harmonic generation with a Ti:Sapphire laser ($\lambda$=780 nm, pulse duration of 120 fs, and repetition rate of 76 MHz). The laser is used to pump the 2 mm type-II $\beta$-barium borate (BBO) crystal to create polarization-entangled photon pairs $\rho_{\phi}$ by the SPDC process. The 1 mm C BBO and C HWP are used to walk-off compensation.}\label{setup}
\end{figure}

The second step is to perform the measurements. When considering the Bell test experiment in which Alice and Bob measure the operators $A_{a}$ and $B_{b}$, according to the original designations \cite{Collins02}, have the nondegenerate eigenvectors
\begin{eqnarray}
&&\left|k\right\rangle_{A,a}=\frac{1}{\sqrt{d}}\sum_{j=0}^{d-1}\exp\big[i \frac{2\pi }{d}j(k+\alpha_{a})\big]\left|j\right\rangle_{A},\nonumber\\
&&\left|l\right\rangle_{B,b}=\frac{1}{\sqrt{d}}\sum_{j=0}^{d-1}\exp\big[i \frac{2\pi }{d}j(-l+\beta_{b})\big]\left|j\right\rangle_{B},\nonumber
\end{eqnarray} 
where $(\alpha_{1},\alpha_{2})=(0,1/2)$ and $(\beta_{1},\beta_{2})=(1/4,-1/4)$. To take measurements from single two-state quantum particles, we rephrase $\left|k\right\rangle_{A,a}$ and $\left|l\right\rangle_{B,b}$ in terms of individual states as
\begin{eqnarray}
\left|k\right\rangle_{A,a}=\bigotimes_{m=1}^{N}\left|k_{m}\right\rangle_{A,a},\left|l\right\rangle_{B,b}=\bigotimes_{m=1}^{N}\left|l_{m}\right\rangle_{B,b}.\label{kl}
\end{eqnarray}
where
\begin{eqnarray}
&&\left|k_{m}\right\rangle_{A,a}=\frac{1}{\sqrt{2}}\big[ \left| 0\right\rangle_{Am} +e^{i \frac{2 \pi}{2^{m}} (k +\alpha_{a})} \left| 1\right\rangle_{Am} \big],\label{km}\\
&&\left|l_{m}\right\rangle_{B,b}=\frac{1}{\sqrt{2}}\big[ \left| 0\right\rangle_{Bm} +e^{i \frac{2 \pi}{2^{m}} (-l+\beta_{b})} \left| 1\right\rangle_{Bm} \big].\label{lm}
\end{eqnarray}
Here we have used the method as the technique for decomposing the quantum Fourier transform into product representation \cite{Nielsen00}. 
For the experimental state $\rho_{\Phi}(d)=\bigotimes_{m=1}^{N}\rho_{\phi m}$ and $\rho_{\phi m}$ denotes the state of the $m$th entangled pair, the joint probabilities can be represented in terms of the probabilities of individual pairs of entangled states by 
\begin{eqnarray}
&&P_{QM}(A_{a}=k,B_{b}=l)\nonumber\\
&&=\text{Tr}[\left|k\right\rangle_{{A,a}\;{A,a}}\left\langle k\right|\otimes \left|l\right\rangle_{{B,b}\;{B,b}}\left\langle l\right|\rho_{\Phi}(d)]\nonumber\\
&&=\text{Tr}\bigg[\bigotimes_{m=1}^{N}\bigg(\left|k_{m}\right\rangle_{A,a\;A,a}\left\langle k_{m}\right|\otimes \left|l_{m}\right\rangle_{B,b\;B,b}\left\langle l_{m}\right|\bigg)\bigotimes_{m'=1}^{N}\rho_{\phi m'}\nonumber\bigg]\\
&& =\prod_{m=1}^{N}Tr[\left|k_{m}\right\rangle_{A,aA,a}\! \left\langle k_{m}\right|\otimes \left|l_{m}\right\rangle_{B,bB,b}\! \left\langle l_{m}\right|\rho_{\phi m}], \label{pqm}
\end{eqnarray}
which shows that the joint probability $P_{QM}(A_{a}=k,B_{b}=l)$ can be derived from the individual joint probabilities of obtaining the measurement outcomes $k_{m}$ and $l_{m}$ on the state $\rho_{\phi m}$. Hence, the Bell expression $I_{d}$ can also be determined from the outcomes of measurements of individual entangled pairs. For the case of two ensembles of perfect entangled states $\left|\Phi\right\rangle$, the Bell inequalities can be maximally violated, that is, $I_{d}(QM)\approx2.970$ for extremely large $d$-dimension \cite{Collins02}.

\begin{figure}[t]
\includegraphics[width=8.5cm]{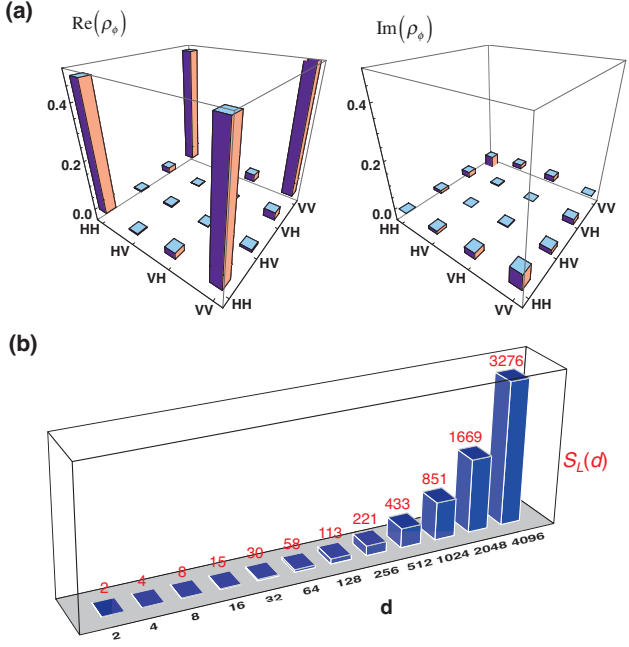}
\caption{Experimental states. (a), The density matrix of experimental state $\rho_{\phi}$. (b), The entanglement dimensionality of the created states $\rho_{\Phi}(d)$ for $d=2,2^{2},...,2^{12}$.}
\label{densitymatrix}
\end{figure}

In the experimental demonstrations of the proposed method, we use entangled photons to construct ensembles $A$ and $B$. As shown in Fig.~\ref{setup}, the ingredient photon pairs are generated through the type-II SPDC process and entangled at the degree of freedom of polarization \cite{Kwiat95,beamlike,beamlike2x2,collinear} which is based on the proposal Fig.~\ref{scheme}. The polarization-entangled pair $\left|\phi\right\rangle=(\left|H\right\rangle_{A}\otimes\left|H\right\rangle_{B}+\left|V\right\rangle_{A}\otimes\left|V\right\rangle_{B})/\sqrt{2}$, where $\left|H(V)\right\rangle$ represents the horizontal (vertical) polarization state. To connect with the conceptual scheme, we make a correspondence of denotations by $\left|H\right\rangle\equiv \left|0\right\rangle$ and $\left|V\right\rangle\equiv\left|1\right\rangle$. In the experiment, the entangled pairs $\rho_{\phi m}$ consisting of the ensembles $A$ and $B$ are created at different times. The stability of our laser and measurement system enables entangled pairs created at different times with a large time separation have a very close fidelity without additional system alignment, which makes the experimental states approximately identical at $\rho_{\phi m}\approx \rho_{\phi}$ for all pairs $m$. The tomographic diagram is depicted in Fig.~\ref{densitymatrix} (a) \cite{James01, Vogel89, Hradil04} and the entanglement source exhibits a high quality by the state fidelity $F_{\phi}=Tr[\rho_{\phi}\left|\phi\right\rangle \left\langle\phi\right|]\approx 0.982 \pm 0.006$.

\begin{figure}[t]
\includegraphics[width=8cm]{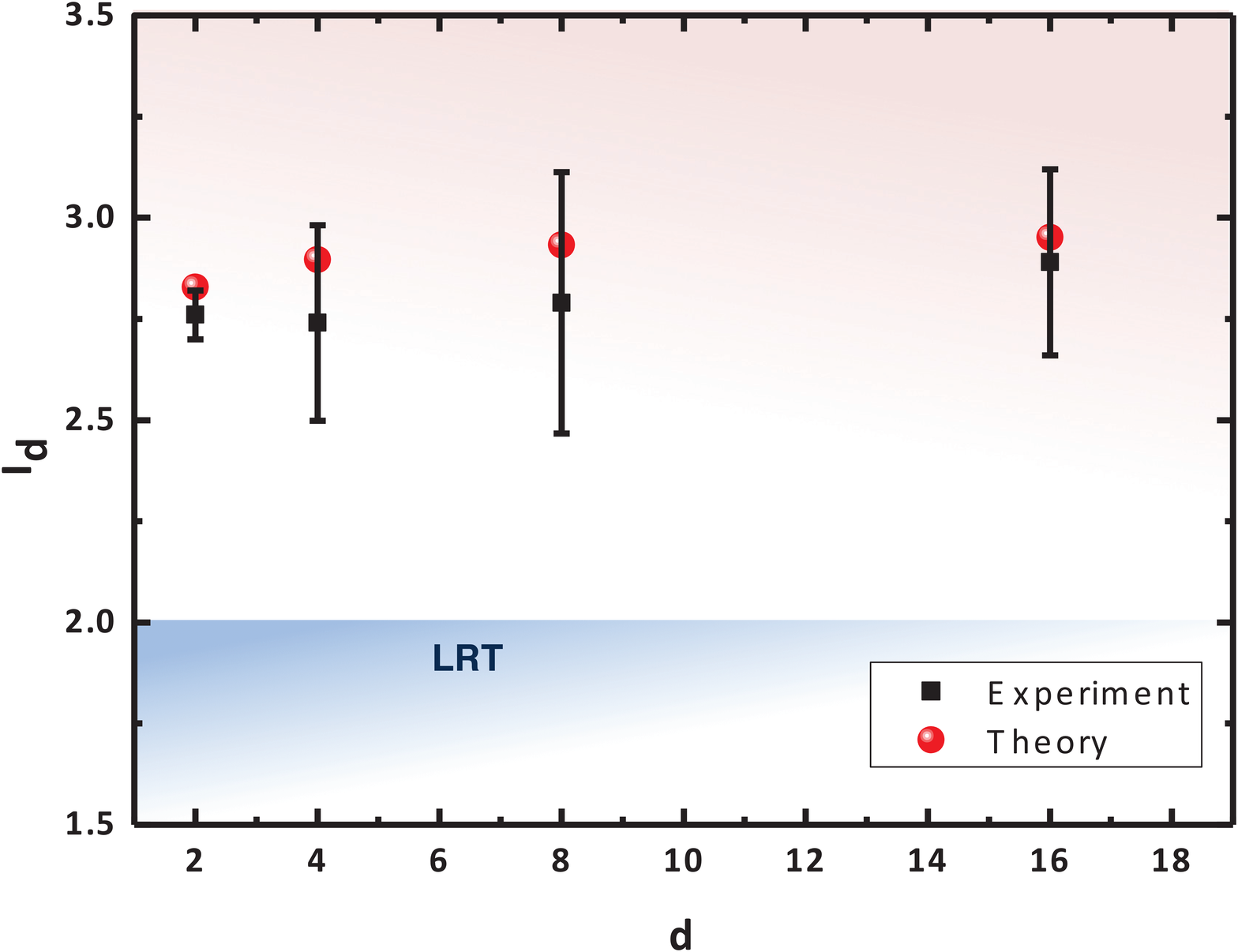}
\caption{Measurement of Bell expressions $I_{d}$. The values of the Bell expressions are determined measuring the probabilities $P_{QM}(A_{a}=k,B_{b}=l)$ [see Eq.~(\ref{pqm})]. ($I_{2}=2.76\pm 0.06$, $I_{4}=2.74\pm 0.24$, $I_{8}=2.79\pm 0.32$ and $I_{16}=2.94\pm 0.46$)}\label{measurementresult}
\end{figure}

\begin{figure}[t]
\includegraphics[width=8cm]{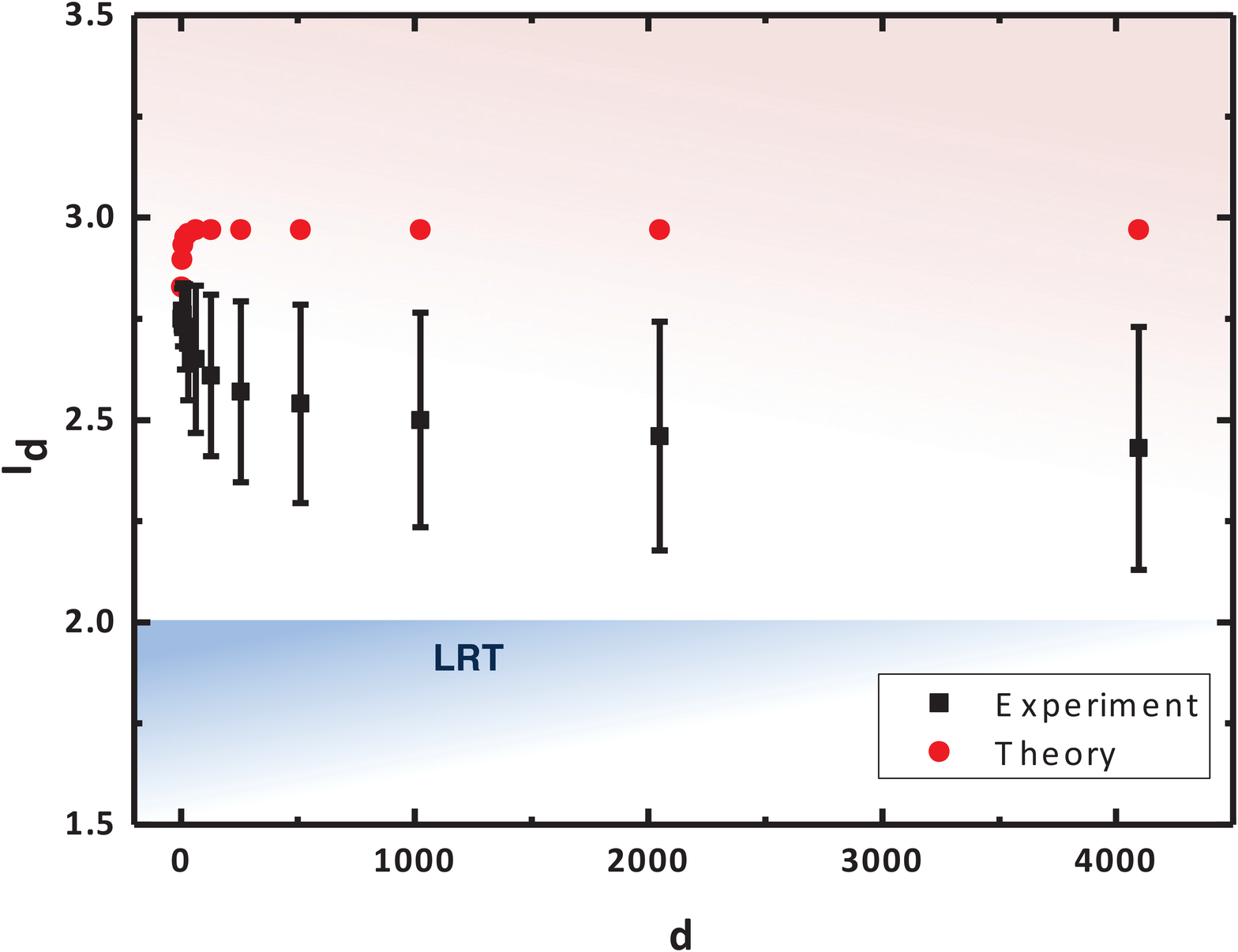}
\caption{Estimations of Bell expressions $I_{d}$. The values of the Bell expressions are estimated by approach measuring the density matrix $\rho_{\Phi}(d)$ \cite{James01}. The results based on this approach agrees with the measured values shown in Fig.~\ref{measurementresult}. 
Violation of the Bell-CGLMP inequality [Eq.~(\ref{belli})] is shown for large $d$ for both of the two approaches overcoming the limit of $I_{d}$ (\ref{kernel}) predicted in LRT. ($I_{2}=2.75\pm 0.03$, $I_{4}=2.77\pm 0.05$, $I_{8}=2.76\pm 0.08$ and $I_{16}=2.73\pm 0.10$) }\label{estimateresult}
\end{figure}

In addition to the Bell-CGLMP inequalities, we use the measured state fidelity $F_{\Phi}(d)$ to extract information about the dimensionality of the entanglement between ensembles $A$ and $B$. Here the entangled dimension is defined in terms of Schmidt number, $S(d)$, of the created state $\rho_{\Phi}(d)$. For an ideal case, the Schmidt number of a perfectly prepared state $\rho_{\Phi}(d)=\left|\Phi\right\rangle \left\langle\Phi\right|$ is $S(d)=d$. We use the Schmidt number witness \cite{Terha00,Sanpera01,Li10} to obtain the lower bounds of the dimensionality of the entanglement. If $\rho_{\Phi}$ shows that $F_{\Phi}(d)=Tr[\rho_{\Phi}(d)\left|\Phi\right\rangle \left\langle\Phi\right|]>(\gamma_{s}-1)/d$ then $\rho_{\Phi}(d)$ is a genuine multi-dimensional entangled state with $S(d)\geq \gamma_{s}$ where $\gamma_{s}\in\{2,3,...,d\}$. For a given state fidelity $F_{\Phi}(d)$, one can find a maximum value of $\gamma_{s}$ which satisfies the above condition, meaning that the maximum lower bound of $S(d)$ can be found as well. Figure~\ref{densitymatrix}(b) illustrates the maximum lower bounds, $S_{L}(d)$, observed in our experiment.

To realize measurements in bases $\{\left|k_{m}\right\rangle_{A,a}\}$ [Eq. (\ref{km})] and $\{\left|l_{m}\right\rangle_{B,b}\}$ [Eq. (\ref{lm})], we use a half-wave plate (HWP) set at $\theta$ and quarter-wave plate (QWP) set at $-\pi/4$ to perform unitary transformations of single-photon polarization states. The angular settings of HWP for the qubits of Alice and Bob are designed as $\theta_{A}=-\pi/8-2\pi(k+\alpha_{a})/4d$ and $\theta_{B}=-\pi/8-2\pi(-l+\beta_{b})/4d$, respectively. See Methods for the complete derivation. After rotating the wave-plate sets, the states of polarizations are projected onto the base $\{\left|H\right\rangle,\left|V\right\rangle\}$.
For a given pair of operators $(A_{a},B_{b})$, the total number of measurement settings of wave plates is $d-1$. Although arbitrary unitary transformations of single-photon polarization states can be performed with high precision by sets of wave plates, the imperfect angle settings introduce errors that accumulate with increasing $d$. Such experimental imperfections become rather crucial when $d$ is large. 

In the experimental demonstration of $d$-dimensional nonlocality, we investigate the Bell expressions using two approaches. As quantum mechanics predicts that the Bell expression follows the definition
\begin{equation}
I_{d}(QM)=Tr[\hat{I}_{d}\rho_{\Phi}(d)].\label{idQM}
\end{equation}
where $\rho_{\Phi}(d)=\rho_{\phi}^{\otimes N}$ and $\hat{I}_{d}$ is the operator of Bell expression, we can obtain the values of $I_{d}(QM)$ either by measuring the density operator of $\rho_{\Phi}(d)$ to calculate $Tr[\hat{I}_{d}\rho_{\Phi}(d)]$ or by performing measurements on the created states, that is, by measuring in the bases of eigenvectors of $A_{a}$ and $B_{b}$ [Eq. (\ref{kl})] to find all of the probabilities $P_{QM}(A_{a}=k,B_{b}=l)$ for $I_{d}(QM)$.

Our experiment shows the violation of the Bell-CGLMP inequalities for systems of $16$ dimensions. The Bell expressions $I_{d}(QM)$ are calculated by measuring all of the probabilities $P_{QM}(A_{a}=k,B_{b}=l)$. As seen in Fig.~\ref{measurementresult}, the experimental results are strongly consistent with the theoretical predictions based on an ideal entangled state $\left|\Phi\right\rangle$ and perfect measurements in Eq.~(\ref{kl}). The Bell expressions measured here are strictly dependent on the accurate settings of the wave plates, as the required setting accuracy increases with $d$ proportionally. Although our demonstration shows cases up to $d=16$ only, the method can be straightforwardly extended to test Bell inequalities for systems of larger dimensionality.

We also made an estimation from the density matrices. Figure~\ref{estimateresult} illustrates that the bipartite $d$-dimensional system composed of polarization-entangled pairs possesses high-quality entanglement  sufficient to violate Bell inequalities for $d>4000$. The density matrix of the created state $\rho_{\Phi}(d)$ was experimentally measured to determine the values of $I_d (QM)$ [Eq. (\ref{idQM})] as the measurement of Bell expression. The degree of quantum violations depends on the purity of $\rho_{\Phi}(d)$ which is evaluated by the state fidelity $F_{\Phi}(d)$. For the stability of our experimental conditions, the state fidelity $F_{\Phi}(d)$ is determined by measuring $F_{\phi}$ under the approximation $F_{\Phi}(d)\approx F_{\phi}^{N}$ which decreases with the increasing number of pairs $N$.


\section*{Discussion}

\begin{figure}
\includegraphics[width=9.5 cm,angle=0]{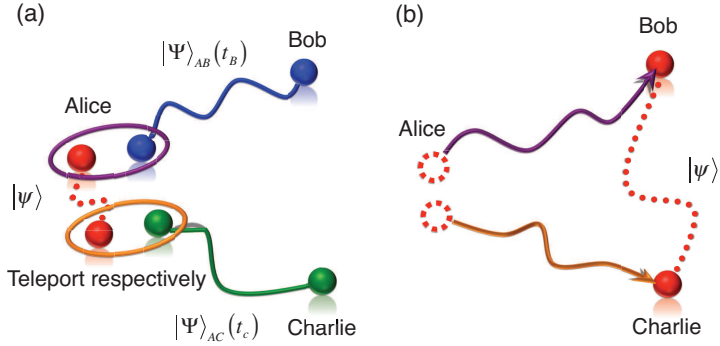}
\caption{Teleportating entanglement with two entangled pairs created at different times. Alice wants to teleport an entangled state $\left|\psi\right\rangle$ to Bob and Charlie. To achieve this aim, as shown in (a), she can use the entangled pair created by Bob, $\left|\Psi\right\rangle_{AB}$ at time $t_{B}$, to teleport partial state of the bipartite entanglement from her hand to Bob's end first. Then, using Charlie's entangled pair, $\left|\Psi\right\rangle_{AC}$ created at time $t_{C}$, where $t_{C}>t_{B}$, Alice performs the subsequent teleportation to complete entanglement sharing between Bob and Charlie (b). While two entangled pairs are created and then used at different times, their quantum nonlocality eventually powers such two-qubit entanglement teleportation. Our scenario can concretely reveal the quantum nonlocality and clearly illustrate such nonclassical source by violating the CGLMP-Bell inequality.}
\label{tel}
\end{figure}

In this experiment, entangled photon pairs are generated to serve as a bipartite $d$-dimensional system to provide stronger non-locality than usual two-level cases. The concept is similar to that two entangled pairs can teleport more than one-qubit information \cite{Zhang06}. For the feature that the Bell-CGLMP inequality can be tested by using entangled photon pairs created at different times, it becomes possible that one can use different experimental output states but have the same state vectors to obtain the same quantum violations. Such situations could be possible in practical quantum information tasks and their applications. For example, Alice wants to use two entangled pairs which are shared with Bob and Charlie respectively to teleport a two-qubit entangled system on her hand such that Bob and Charlie jointly share this two-qubit entanglement. In general, these two entangled pairs can be created with different entanglement sources and then used for teleportation at different times due to some technical reasons, for example, the mediate entangled pairs are generated by Bob and Charlie separately. See Fig.~\ref{tel}. This can be considered as an extension of the scenario used in the experiment of Zhang \textit{et al.} \cite{Zhang06} where the two entangled pairs are created at the same time. While the entangled pairs are created at different times by using different sources, the target two-qubit entangled state is enabled to be teleported in principle. Our scenario can concretely reveal the quantum nonlocality that powers this two-qubit teleportation by violating the Bell-CGLMP inequality.
 
Extending this idea, a multi-partite multi-dimensional system will be implemented in the future work by genuine multi-partite entanglement of two-dimensional systems. With this approach, it would be possible to investigate generic quantum nonlocality by systematically implementing Bell test experiments for the multi-partite multi-dimensional systems \cite{Son06}. There have been extensive investigations of multi-partite entanglement, both from the fundamental and practical points of view. In particular, a genuine eight-photon polarization-entangled state has recently been experimentally generated using the SPDC process \cite{Yao12}. Hence, our novel scheme holds high promise for testing generic Bell inequalities for the eight-partite high-dimensional systems using a state-of-the-art eight-photon entanglement source. In addition to entangled photons, one can directly apply our idea to other quantum systems where the states are able to be coherently manipulated and locally measured. Such kind of Bell test experiments could be realized in systems of multi-partite entangled ions in the Greenberger-Horne-Zeilinger (GHZ) state \cite{Monz11}. The systems of nanomechanical resonators \cite{Johansson14} and Josephson circuits \cite{Wei05,Wei06} also provide these suitable conditions to implement our scheme. Furthermore, it is interesting to investigate violations of CGLMP-Bell inequality with multi-dimensional entanglement composed of intra- and inter-pair entangled states \cite{Ashhab09}. While the devised scenario still requires a rigorous theory showing its general utility in quantum applications such as quantum information processing, it provides a novel way to consider all the related information tasks based on the generic Bell test experiments.  

Finally, we remark three points on the present experimental violations of Bell-CGLMP inequalities. First, while the ensemble of photon pairs is mathematically equivalent to a multi-dimensional entangled state, each Bell pair is generated and immediately afterwards already detected, and then such a multi-dimensional entangled state is not generated at any point of time. Here, our aim is to demonstrate that the test of Bell-CGLMP inequalities can be shown by sub-systems even though the particles never co-existed. Second, using the independently generated Bell pairs to realize a quantum information protocol, such as quantum teleportation introduced above (Fig.~\ref{tel}), can use quantum memory\cite{Chen08,Yuan10} for state storage so that they can co-exist before the subsequent required measurements. This is different from the implementation which utilizes multi-dimensional states of genuinely single quantum systems \cite{Dada11}. Compared with the teleportation scheme without memory, Fig.~\ref{tel}, the states $\left| \psi\right\rangle$ can be teleported to Bob and Charlie at the same time with the support of quantum memory, which gives higher flexibility for a subsequent quantum information task. Finally, in addition to the present experimental demonstration using SPDC source, there exists other potential methods to create ensembles of entangled photons to show the general utility of our idea in the Bell test experiments on multi-dimensional systems. For instance, one can use hyperentangled states to investigate quantum violations of Bell-CGLMP inequalities. As shown in the experiments on such entanglement \cite{Yang05,Chen07}, the two-photon four-qubit state entangled both in polarization and spatial modes has been created as $1/\sqrt{2}(\left|H\right\rangle_{A}\left|H\right\rangle_{B}+\left|V\right\rangle_{A}\left|V\right\rangle_{B})\otimes1/\sqrt{2}(\left|R\right\rangle_{A}\left|R\right\rangle_{B}+\left|L\right\rangle_{A}\left|L\right\rangle_{B})$, where $\left|R\right\rangle_{A(B)}$ and $\left|L\right\rangle_{A(B)}$ represent two orthogonal states with the spatial modes $R_{A(B)}$ and $L_{A(B)}$. This hyperentangled state is equivalent to a maximally entangled state of two four-dimensional systems [See Eq.~(\ref{Phi})] and then suitable for implementing our scheme.

In conclusion, we have demonstrated the violation of the Bell-CGLMP inequalities for systems of extremely large dimensions. These demonstrations are based on bipartite $d$-dimensional systems composed of polarization-entangled photon pairs. Using the well-known technique of type-II SPDC for polarization entanglement, which is robust and stable and needs only modest experimental effort using standard technical devices such as wave-plate sets and SPCMs, the scheme provides a way to experimentally observe $d$-dimensional quantum nonlocality up to $d=16$ by performing measurements of the Bell expressions $I_{d}$. An estimation of the quantum nonlocality also shows up to $d=4096$ by measuring the density matrix of the entangled state.

\section*{Methods}

\noindent \textbf{Measurements on single entangled photon pairs}

\noindent To realize the measurements according to Eq.~(\ref{pqm}), the probabilities are determined by firstly performing local transformation on photons. We use HWP and QWP to perform the required rotations of polarization states.  See Fig.~\ref{setup}. The HWP and QWP can be described by the following transformations, respectively \cite{James01}:
\begin{eqnarray}
&&H(\theta)=\cos(2\theta)(\left|H\right\rangle\!\!\left\langle H\right|-\left|V\right\rangle\!\!\left\langle V\right|)-\sin(2\theta)(\left|H\right\rangle\!\!\left\langle V\right|+\left|V\right\rangle\!\!\left\langle H\right|),\nonumber\\
&&Q(\gamma)=[i-\cos(2\gamma)]\left|H\right\rangle\!\!\left\langle H\right|+[i+\cos(2\gamma)]\left|V\right\rangle\!\!\left\langle V\right|+\sin(2\gamma)(\left|H\right\rangle\!\!\left\langle V\right|+\left|V\right\rangle\!\!\left\langle H\right|),
\end{eqnarray} 
where $\left|H(V)\right\rangle$ denotes the photon state with horizontal (vertical) polarization. Hence, with combination of both wave plates, the local transformation for polarization state is 
\begin{eqnarray}
U(\theta)&=&H(\theta)Q(-\frac{\pi}{4})\nonumber\\
&=&-\left|H\right\rangle\frac{1}{\sqrt{2}}(\left\langle H\right|+e^{i(4\theta+\pi/2)}\left\langle V\right|)+i\left| V\right\rangle\frac{1}{\sqrt{2}}(\left\langle H\right|-e^{i(4\theta+\pi/2)}\left\langle V\right|).
\label{HQ}
\end{eqnarray}
To implement measurements onto the state sets $\{\left| k_{m}\right\rangle_{A,a}\}$ and $\{\left|l_{m}\right\rangle_{B,b}\}$, the angular settings of HWP for the qubits of Alice and Bob are designed as
\begin{eqnarray}
\theta_{A}&=&-\pi/8-2\pi (k+\alpha_{a})/4d,  \\
\theta_{B}&=&-\pi/8-2\pi (-l+\beta_{b})/4d,
\end{eqnarray}
respectively. After such transformations, the states $\left| k_{m}\right\rangle_{A,a}$ and $\left| l_{m}\right\rangle_{B,b}$ will become horizontal-polarization states $\left|H\right\rangle_{A}$ and $\left|H\right\rangle_{B}$. Thus determining the joint probability of measuring $\left| k_{m}\right\rangle_{A,a}$ and $\left|l_{m}\right\rangle_{B,b}$ for $P(A_{a},B_{b})$ is equivalent to measuring the joint probability $P(H,H)$ on the $m$-th entangled photon pair $[U(\theta_{A})\otimes U(\theta_{B})]\rho_{\phi m}[U(\theta_{A})\otimes U(\theta_{B})]^{\dag}$.\\

\section*{Acknowledgements}
This work was supported by the Ministry of Science and Technology, Taiwan, Republic of China (Contract No. 101-2112-M-009-002-MY2, 103-2112-M-009-008, 101-2112-M-006-016-MY3, 101-2628-M-006-003-MY3, 101-2112-M-009-016-MY2, 103-2119-M-009-004-MY3, 103-2628-M-009-002-MY3, 103-2923-M-009-001-MY3, 103-2112-M-006-017-MY4, 104-2112-M-009-001-MY2, and 104-2112-M-006-016-MY3), and the Grant MOE ATU Program at NCTU.

\section*{Author Contributions}
H.P.L., A.Y., C.W.L. and T.K. carried out the experiment; C.-M.L. devised the theoretical and experimental scheme and Y.-N.C. provided the theoretical analysis; H.P.L., A.Y., C.W.L. and T.K. conceived the experiment and analysed the data; and all authors co-wrote the paper. 

\section*{Additional Information}
{\bf Competing financial interests:} The authors declare no competing financial interests.

\end{document}